# Uncovering the hidden order in URu$_2$Si$_2$: Identification of Fermi surface instability and gapping

S. Elgazzar[1], J. Rusz[1], M. Amft[1], P. M. Oppeneer[1]*, J.A. Mydosh[2]

[1] *Department of Physics and Materials Science, Uppsala University,*
   *Box 530, S-751 21 Uppsala, Sweden*

[2] *II. Physikalisches Institut, Universität zu Köln, Zülpicher Strasse 77, D-50937 Köln, Germany*

**Spontaneous, collective ordering of electronic degrees of freedom leads to second-order phase transitions that are characterized by an order parameter. The notion "hidden order"(HO) has recently been used for a variety of materials where a clear phase transition occurs to a phase *without* a known order parameter. The prototype example is the heavy-fermion compound URu$_2$Si$_2$ where a mysterious HO transition occurs at 17.5K. For more than twenty years this system has been studied theoretically and experimentally without a firm grasp of the underlying physics. Using state-of-the-art density-functional theory calculations, we provide here a microscopic explanation for the HO. We identify the Fermi surface "hot spots" where degeneracy induces a Fermi surface instability and quantify how symmetry breaking lifts the degeneracy, causing a surprisingly large Fermi surface gapping. As mechanism for the HO we propose spontaneous symmetry breaking through collective antiferromagnetic moment excitations.**

*E-mail: peter.oppeneer@fysik.uu.se

The hidden order (HO) state in the uranium compound $URu_2Si_2$ has had a mystifying attraction since its discovery 20 years ago [1-3]. As temperature is reduced below 100K, lattice coherence between local uranium $f$-moments in $URu_2Si_2$ develops and a heavy-fermion liquid starts to form as the uranium moments are dissolved into the Fermi surface (FS). At $T_0$=17.5K the HO transition takes place, evidenced by dramatic effects in the thermodynamic and transport properties [1-3]. While the latter properties unambiguously exhibit sharp anomalies, typical of a magnetic phase transition, microscopic measurements, such as neutron diffraction, μSR and nuclear magnetic resonance (NMR), do not indicate a transition to a well-ordered magnetic phase. Very small antiferromagnetic (AF) moments of ~0.03$μ_B$ have been detected in the HO [4], which by far are too small to explain the large entropy loss and sharp anomalies in the thermodynamic quantities [5,6]. However, with pressure the U-moments are resurrected (~0.4$μ_B$) and an ordered, large moment antiferromagnetic (LMAF) state develops. The bulk properties between HO and LMAF are very much alike with similar, continuous changes in the thermodynamic quantities. Due to this equivalence, the term "*adiabatic continuity*" was adopted to describe the pressure transformation from HO to LMAF and its corresponding physics [7]. Other recent measurements indicate a mild change in slope of pressure-dependent quantities at the HO to LMAF crossover [8,9]. Below 1.5K and only out of the HO an exotic superconducting state appears, which is the subject of recent interest [10,11].

A variety of theories have been proposed to explain the mysterious order parameter of the HO state [12-18]. Some of these theories assume localized uranium $f$-states, others itinerant $f$-states. Yet, there is no general agreement as to a model to fully describe the HO. Neutron scattering measurements [19,20] showed that time-reversal symmetry must be broken in both the HO and LMAF phase. Experimental investigations such as conductivity measurements [3,8], Hall effect [21], infrared spectroscopy [22] and thermal transport [23] are consistent with the opening of a gap at the Fermi surface in both the HO and LMAF phases. Compatibly, recent inelastic neutron scattering experiments revealed gapped spin excitations in the HO phase [24]. Since the high-temperature paramagnetic (PM) phase possesses a non-gapped FS, the anticipated marked changes in FS topology at the HO/AF transition, though occurring at a small energy scale of about 10meV, should be traceable by band-structure calculations. A few such calculations have been performed for $URu_2Si_2$ [25,26]. However, these earlier investigations were not yet accurate enough to provide a clear picture neither of the energy dispersions nor the Fermi surface, and in particular did not attempt to compare the FS for the different phases. A materials-specific, microscopic model of how and where the FS gapping occurs is consequently lacking.

We have performed density-functional theory (DFT) investigations of the energy band dispersions and Fermi surface of $URu_2Si_2$, using two state-of-the-art electronic structure methods (see Methods section for details). We tested that these two independent full-potential relativistic codes provide *precisely* the same energy dispersions and FS. Our calculations have been done for both the PM and the LMAF state; our aim is to explain these two phases first and subsequently converge on the HO phase. The large-moment (type-I) AF phase has a computed uranium magnetic moment of $0.39\mu_B$ ($-0.36\mu_B$ spin and $0.75\mu_B$ orbital moment, along the *c*-axis of the $ThCr_2Si_2$ unit cell), which compares very well to the experimental moment of $0.40\mu_B$ [9]. Also, we have investigated the development of a continuous transition from the PM to LMAF phase by varying the exchange interaction, responsible for the formation of the magnetic moment. In this manner we can gradually transform from a small-moment antiferromagnetic (SMAF) state with total moment of $\sim 0.03\mu_B$ to the LMAF state. We emphasize that while the SMAF and HO phases do have similarities, we do not state that they are identical, but rather use the SMAF phase to study the consequences of the small moment.

**Figure 1** shows the computed electronic bands for $URu_2Si_2$ in both the PM and LMAF phases. For a convenient comparison, both sets of energy dispersions are shown in the tetragonal Brillouin zone (BZ) of the LMAF phase. Figure 1 evidences that the difference between the electronic structures of the two phases is quite small: the energy dispersions of the two phases are practically superimposed. Some bands, which are degenerate in the PM phase, become split in the LMAF phase, due to its exchange interaction (e.g., along the Z-A and R-Z high-symmetry directions). At the Fermi energy ($E_F$), which is at 0 eV, the PM and AF dispersions are mainly coinciding throughout the whole BZ, *except* for one conspicuous point along the Γ-M (Σ) high-symmetry line and at the Z point. The Z-point modification is insignificant, as will become clear below. However, the modification along the Γ-M direction is indeed significant: a crossing of two energy bands occurs, causing a band degeneracy in the PM phase which osculates the Fermi level. For clarity this reciprocal space section is highlighted by the green box and shown enlarged in the inset of Fig. 1. AF ordering provides a symmetry breaking that removes the band degeneracy and causes the opening of a gap at the Fermi level. The same two PM band dispersions also cross close to $E_F$ at other, non high-symmetry points in the BZ. They additionally cross along the X-Γ (Δ) high-symmetry line but there AF ordering does not lift the degeneracy.

**Figure 2** shows the computed FSs for the PM and LMAF phases. The FSs we have computed are significantly different from the earlier ones [25,26]. We obtain for both phases three (Kramers doubly-degenerate) bands that cross the Fermi level. Panels a-c display the resulting three FS sheets in the PM phase while panels d, e, and c show the corresponding sheets for the LMAF phase. The third FS sheet (panel c) is the same in both phases and therefore displayed only once. A comparison of the FS sheets in a and d, and b and e, respectively, exemplifies how dramatic the modification of the FS due to the removal of the degeneracy is. In Fig. 2a there is a rugged, four-armed FS sheet that becomes *completely* gapped in the LMAF phase (Fig. 2d). The same band gives rise to a Γ-centered rugby-ball shaped sheet, which however is identical in both phases. Panels b and e show that the second FS sheet is also modified: there are four cup-like FS parts in the PM phase which shrink substantially in the LMAF phase (Fig. 2e). The above-mentioned difference between the PM and LMAF energy bands at the Z point leads to an insignificant FS change, as can be recognized from panels b and e. Panel f shows a cross-section of the two crucial FS sheets in the $k_z$=0 plane. Blue symbols depict the computed FS contours for the PM, red symbols for the LMAF phase. The rugged, arm-shaped FS sheets in Fig. 2a correspond to the blue ellipses centered on the Γ-M high-symmetry lines whereas the half-sphere FS sheets of LMAF URu$_2$Si$_2$ shown in Fig. 2e relate to the red contours centered on the Γ-X lines. Note that the PM FS consists, in fact, of two intersecting sheets. A degenerate crossing point of the PM bands occurs precisely at $E_F$ at the intersection lines of these two sheets, i.e., forming Dirac-like crossings between the Γ-M and Γ-X directions in the $k_z$=0 plane (indicated by the purple arrow in Fig. 2f). Degenerate crossings occur in the whole BZ on eight intersection lines of the two PM FS sheets, which extend almost from the Z to –Z points.

We have investigated how the gapping commences at the degenerate crossing points by performing calculations for a range of AF phases, from the SMAF phase up to the LMAF phase, through the gradual variation of the exchange interaction. Our calculations for the SMAF phase reveal that already for tiny moments of 0.03$\mu_B$, the entangled FS sheets break-up at the intersection lines and form disconnected, smaller FS sheets. The induced FS gap is then smaller than that occurring for the LMAF phase. As shown below, the gapping increases (nonlinearly) with increasing moment and, accordingly, the arm-like FS sheets shrink until these become completely gapped when the full-moment of 0.39$\mu_B$ is nearly reached. While these calculations have been performed for various AF phases, we conjecture that we have identified the FS "hot spots" for the HO transition. Such critical regions on the FS are where the gapping takes place and the size of which is related to the order parameter of the HO.

**Figure 3** quantifies how the FS gap progresses with the total moment per uranium atom. The total U-moment along the *c*-axis is orbitally-dominated, with the orbital moment being twice larger than the spin moment. The FS gap value depends on the position in *k*-space. For small U total moments (0.04$\mu_B$) a gap of about 7meV first occurs at the FS hot spots. When the moment increases, the gap increases and the position of the maximal gap in *k*-space shifts from the hot spot to the center of the ellipse at the Γ-M line. The maximum gap is reached along the Γ-M direction for the LMAF phase.

Next, we show that our calculations are fully consistent with the known experimental properties of URu$_2$Si$_2$. All our calculations demonstrate that the differences between the PM and LMAF phases are extremely small. The LMAF total energy is computed to be only 7K per formula unit deeper than the PM total energy. We have computed the theoretical equilibrium volume, which is only 1.6% smaller than the experimental volume. A recent experimental work [10] classified HO URu$_2$Si$_2$ to be a low-carrier density, electron-hole compensated metal, which is indeed the case for the computed electronic structure as a closer inspection of the band dispersions in Fig. 1 demonstrates. From the computed plasma frequency we obtain for the number of holes per uranium atom $n_h$=0.0185 in the LMAF phase, which compares favorable with the recent values 0.017<$n_h$<0.021 [10]. The number of holes of the PM phase is calculated to be four times larger (0.08 per U-atom). Hall effect measurements estimated 0.1 holes per U-atom [10,21] for the PM phase, in reasonable agreement. An analysis of the character of the energy dispersions crossing at $E_F$ reveals that these consist dominantly of itinerant *f*-states, i.e., the gapping occurs as a rearrangement of itinerant *f*-electrons. The itinerant character of the *f*-states involved in the FS transition is consistent with recent neutron scattering experiments [24] that detected itinerant excitations above $T_0$. Dispersive variations of the 5*f*-spectral weight have, furthermore, been observed with angular resolved photoemission [27,28]. The FS cross-sections shown in Fig. 2f contain several quasi-parallel FS sections, stipulating significant nesting. The dominant nesting occurs between the flat side of the cup-like sheet and the opposite-side of the Γ-centered squarish FS sheet, with nesting wavevector of $0.4a^*$ ($a^* = 2\pi/a$). Neutron scattering experiments [24,29] identified nesting at wavevectors $Q = (1-0.4, 0, 0)$ and $(1+0.4, 0, 0)$ in perfect agreement.

Maple et al. [3] originally proposed a gapping of the FS at the transition to the HO/AF from transport measurements; the BZ-averaged gap was approximated to be Δ ~11meV [3]. Somewhat smaller gaps Δ ~5–7meV were obtained by Mentink et al. [30] and Jeffries et al. [8]. The gap we have identified here varies from maximally 65meV along the Σ-direction to 0meV along the Δ-direction for the LMAF state. For the SMAF phase the gap opening at the intersection lines is about 7meV. Our computed gaps are larger than the experimental (transport) one because they are $k$-dependent and derived for the coherent ($T$=0K) phases, whereas the experimental analysis assumes a $T$-independent gap from 0K to $T_0$. We have furthermore determined the effect of the FS gapping on the resistivity by computing the conductivities in the Kubo formalism. The theoretical resistivity change due to the removal of the FS instability is large and anisotropic, $[\rho_{AF} - \rho_{PM}]/\rho_{PM} (J \| c) = 620\%$ and $[\rho_{AF} - \rho_{PM}]/\rho_{PM} (J \| a) = 160\%$, for the LMAF phase. These values are larger than the experimental values of 56% and 14% at the HO transition; however, the experimental resistivity is superimposed on a large $T^2$-dependent background (due to incoherent and phonon scattering), which is *not* included in our ($T$=0K) calculations. With background subtracted, the experimental values would be about 400% and 100%, respectively. A considerable anisotropy of $[\rho_{HO} - \rho_{PM}]/\rho_{PM}$ has been observed experimentally [31] in agreement with our theoretical ratio of 4:1.

Bonn et al. [22] performed infrared optical spectroscopy on URu$_2$Si$_2$ and observed a reduction of its optical conductivity $\sigma(\omega)$ when temperature was lowered below $T_0$, evidencing the opening of a gap in the *a-a* plane. We have calculated the optical conductivity for a range of magnetic moments, from the PM phase up to the LMAF phase. **Figure 4** shows the computed conductivity spectra Re[$\sigma_a(\omega)$] (i.e., $E \| a$) and Re[$\sigma_c(\omega)$] ($E \| c$). For small photon energies (below 50meV) the optical conductivity, being maximal for the PM phase, becomes clearly reduced for the SMAF and LMAF phases, due to the gapping occurring in the vicinity of $E_F$. The largest drop in $\sigma(\omega)$ is obtained for the LMAF phase, where the FS gapping is the largest. The computed behavior agrees with experiment: Bonn et al. [22], who only measured in the *a-a* plane ($E \| a$), observed a reduction of the reflectivity in the HO phase for energies below 30meV. Our calculation predicts a drop in $\sigma_a(\omega)$ for the small-moment state below 40meV (red and blue curves in Fig. 4). Bonn et al. did not measure $\sigma_c(\omega)$ (i.e., $E \| c$), but our theory predicts that a *larger* reduction of $\sigma_c(\omega)$ should occur at small energies below 50meV as well as an inverted behavior, i.e., an increase, at energies of 50meV up to 600meV.

Our calculations are fully consistent with the known experimental properties of URu$_2$Si$_2$ and provide a natural explanation for the reported "adiabatic continuity" [7], i.e., the similarity as well as continuous transition in the bulk physical properties of HO/SMAF and LMAF URu$_2$Si$_2$, which can be understood as an increase in the FS gapping. When we compare the symmetry operations of the body-centered tetragonal PM phase to those of the SMAF and LMAF phases, we evidently obtain that in the latter phases time-reversal symmetry is broken, as there exist small spin and (two-times larger) orbital moments on U. The crystal symmetry remains tetragonal, but, as the two U atoms in the cell are not equivalent, the body-centering is also broken. Both symmetry reductions agree with experimental observations [4,19,20]. According to our band-structure model the amount of FS gapping is related to the exchange splitting (that drives the net longitudinal U magnetic moment along *c*). In most materials a small moment will normally not induce a considerable FS gapping. However, URu$_2$Si$_2$ is a special material, as there exist degenerate band crossings located precisely at $E_F$, the identified FS hot spots. Yet, this does not imply that the HO and SMAF phases are identical. As the observed magnetic moment of the HO phase is tiny [4] or even extrinsic (e.g., due to stress inhomogeneity) [9], we emphasize that the dominant mechanism leading to FS gapping is the combined time-reversal and translational symmetry breaking. When temperature is lowered towards $T_0$ and the *f*-electrons become incorporated in the FS, the degenerate band crossings at $E_F$ emerge, but now the system can effectively remove these degenerate crossings by spontaneous symmetry breaking. As our calculations showed, this could be achieved by a small dipolar moment along the *c*-axis, which would already be sufficient to lift the degeneracy at the FS hot spots and cause a sizable gapping. Nonetheless, in the limit of vanishing spin and orbital moments the FS gapping will vanish as well, whereas experiment indicates that the gapping in the HO phase is only about 20% smaller than in the LMAF [8]. Consequently, spurious AF moments alone are not sufficient to explain the HO and another mechanism that breaks time-reversal and translational symmetry must be responsible for the more robust gapping appearing in the HO.

Since the spontaneous symmetry breaking happens on a very small energy scale (~7K) we must search for a novel mechanism causing the HO transition. An intriguing possibility here would be spontaneous symmetry breaking through a collective mode of long-lived antiferromagnetic moment excitations. If such a mode would be lattice coherent and involve sufficiently large longitudinal AF moments, then our calculations evidence the induced FS gapping to be larger than that of static SMAF order. We note that, since the FS gap is an *even function* of the longitudinal AF moment, the macroscopic gap value, which is obtained through averaging over many lattice sites, does not average out to zero, whereas the macroscopic net moment does average to zero. Given that URu$_2$Si$_2$ is exceptional, in which a substantial FS gapping of 750K occurs through symmetry breaking on a scale less than 7K, such a lattice coherent dynamic AF mode–which otherwise would only be a soft perturbation–can now alter the electronic structure sufficiently, so that the material's bulk properties become modified. The existence of a dynamic AF mode is in accord with several recent experiments. Inelastic neutron scattering studies discovered long-lived modes of longitudinal spin excitations in the HO phase [4,24]. Other inelastic neutron studies observed a sharp excitation at the AF wavevector in the HO phase. Under pressure this dynamic AF mode "freezes" into the static AF order of the LMAF phase [29], i.e., it becomes the static magnetic Bragg peak of the LMAF phase. Also, resonant X-ray scattering revealed an anomalous Bragg peak at the AF wavevector that was associated with a dynamic AF state in the HO phase [32]. The occurrence of the dynamic AF mode additionally provides a scenario for why exotic superconductivity emerges only in the HO and not in the LMAF phase [11]. The coherent AF mode appears as the ideal candidate for mediating the unconventional superconductivity in URu$_2$Si$_2$.

Our calculations provide a microscopic picture for where the Fermi surface instability occurs and how the FS gapping evolves. The order parameter, which follows from our band-structure model, is the instantaneous longitudinal AF moment that determines the magnitude of the FS gap. Any future complete theory of the HO must take this FS instability and gapping into account. The identified FS hot spots could be probed by future high-resolution angular resolved photoemission experiments, and neutron experiments or X-ray magnetic circular dichroism could be exploited to detect the orbitally-dominated magnetism. The mechanism that we propose for the HO is dynamic symmetry breaking through lattice coherent modes of longitudinal AF moment excitations.

**METHODS**

The density-functional theory (DFT) calculations presented here for URu$_2$Si$_2$ were performed using two state-of-the-art relativistic electronic structure codes, the full-potential local orbitals (FPLO) method [33] as well as the full-potential linearized augmented plane wave (FPLAPW) method in the WIEN2k implementation [34]. We employed for the exchange-correlation functional both the local spin-density approximation (LSDA) in the Perdew-Wang parameterization [35] and the generalized gradient approximation (GGA) in the parameterization of Perdew et al. [36]. The $f$-electrons were treated as itinerant valence states and allowed to hybridize with other valence states. Specifically, we used in the FPLO calculations the following sets of basis orbitals: 5$f$;6$s$6$p$6$d$;7$s$7$p$ for U, 4$s$4$p$4$d$;5$s$5$p$, and 3$s$3$p$3$d$, for Ru and Si, respectively. For the FPLAPW calculations we have included relativistic local orbitals in the basis. The computational parameter determining the basis size, $RK_{max}$, was equal to 7.5 in our calculations. We have verified that both electronic structure codes provide the same electronic structure. 20x20x20 $k$-points in the whole BZ were used to resolve the band dispersions close to $E_F$.

**Acknowledgements**
The authors would like to acknowledge helpful discussions with R. Caciuffo, M. Biasini, G.H. Lander, and N. Bernhoeft.
This work was supported by the Swedish Research Council (VR), the Swedish National Infrastructure for Computing (SNIC), STINT, and the European Commission, JRC-ITU.

**Competing financial interests**
The authors declare that they have no competing financial interests.

**Figure 1  The theoretical energy dispersions of URu$_2$Si$_2$.** Energy dispersions of paramagnetic (PM) URu$_2$Si$_2$ are shown in blue, those of large-moment antiferromagnetic (LMAF) URu$_2$Si$_2$ in red. For comparison both set of energy bands are shown for high-symmetry directions in the tetragonal Brillouin zone.  Close to the Fermi energy (at 0 eV) the energy dispersions of the two phases are nearly indistinguishable, except for the bands crossing the Fermi level between the Γ and M points along the Σ high-symmetry direction. The green rectangle highlights the lifting of a degeneracy, which is shown enlarged in the inset: the crossing of two bands close to the Fermi energy in the PM phase is removed in the LMAF phase.

**Figure 2  The Fermi surface of URu$_2$Si$_2$ in the paramagnetic (PM) and large-moment antiferromagnetic (LMAF) phase.** Panels **a** and **b** show the PM Fermi surface sheets, while panels **d** and **e** show the corresponding LMAF sheets. The FS sheet shown in panel **c** is identical for both phases.  The high-symmetry points and directions in reciprocal space are given in this panel. The lifting of the degeneracy in the PM bands crossing at $E_F$ causes the removal of a sizable FS parts: the rugged, four-armed FS bands in **a** are completely gapped in the symmetry-broken phase (panel **d**). The cup-like FS sheets in **b** furthermore shrink substantially (cf. panel **e**). Panel **f** shows a cross-section of the two main FSs in the $k_z=0$ plane; the FS contour of URu$_2$Si$_2$ in the PM phase is given by the blue symbols and that of the LMAF phase by the red symbols. The difference between these two FS contours exemplifies the surprisingly large FS part that is removed by lifting of the degeneracy. The purple arrow indicates the position of one of the FS hot spots, while the green arrow specifies the nesting vector of length 0.4$a^*$.

**Figure 3 Fermi surface gap of URu$_2$Si$_2$ versus the uranium total moment along the *c*-axis.** The gap Δ depends on the position in reciprocal space. Shown are the gap values at the FS hot spots (squares) in the $k_z=0$ plane and the maximal gap values for all *k*-points in the $k_z=0$ plane (circles).

**Figure 4 Calculated optical conductivity Re[$\sigma_a$] (full lines) and Re[$\sigma_c$] (dashed lines) of URu$_2$Si$_2$.** The spectra have been computed for different uranium total magnetic moments, from the PM phase (0.0 μ$_B$), relatively small moment AF (0.09 μ$_B$), to the LMAF phase (0.39 μ$_B$). The effect of the increased gapping in the vicinity of the Fermi energy causes the progressive reduction of $\sigma_a$ and $\sigma_c$ below 50 meV photon energy. The inset shows the computed spectra on a larger energy interval, exemplifying that the influence of the gapping is restricted to small photon energies (up to 600 meV for $\sigma_c$).  For sake of visibility the Re[$\sigma_c$] spectra have been shifted upward by 3 10$^{15}$ s$^{-1}$.

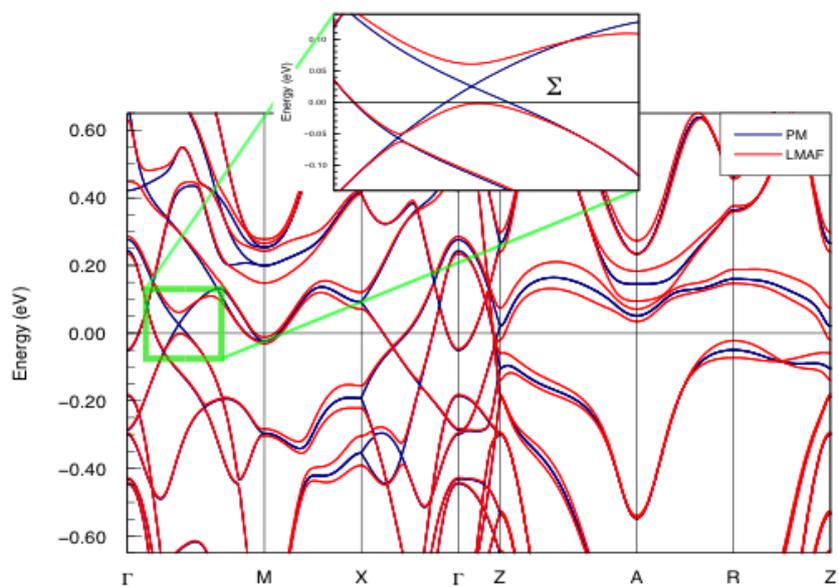

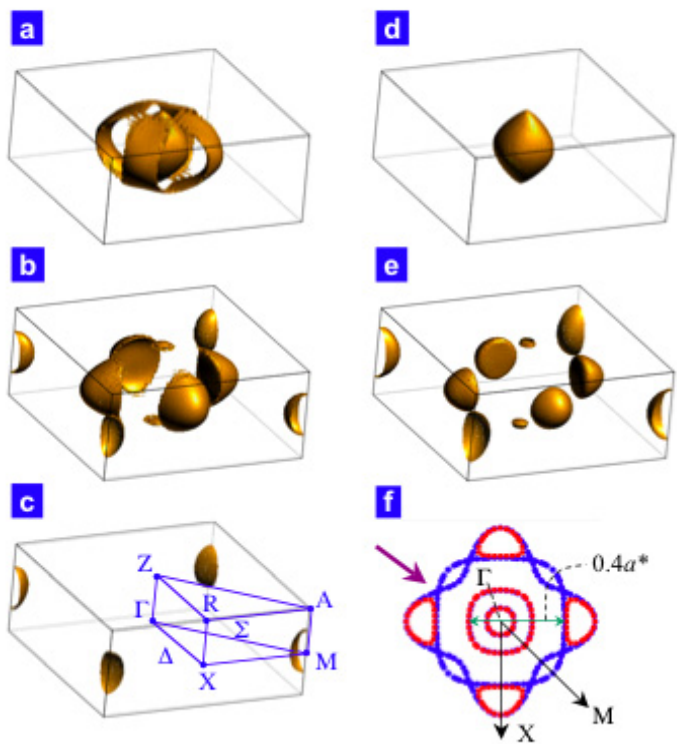

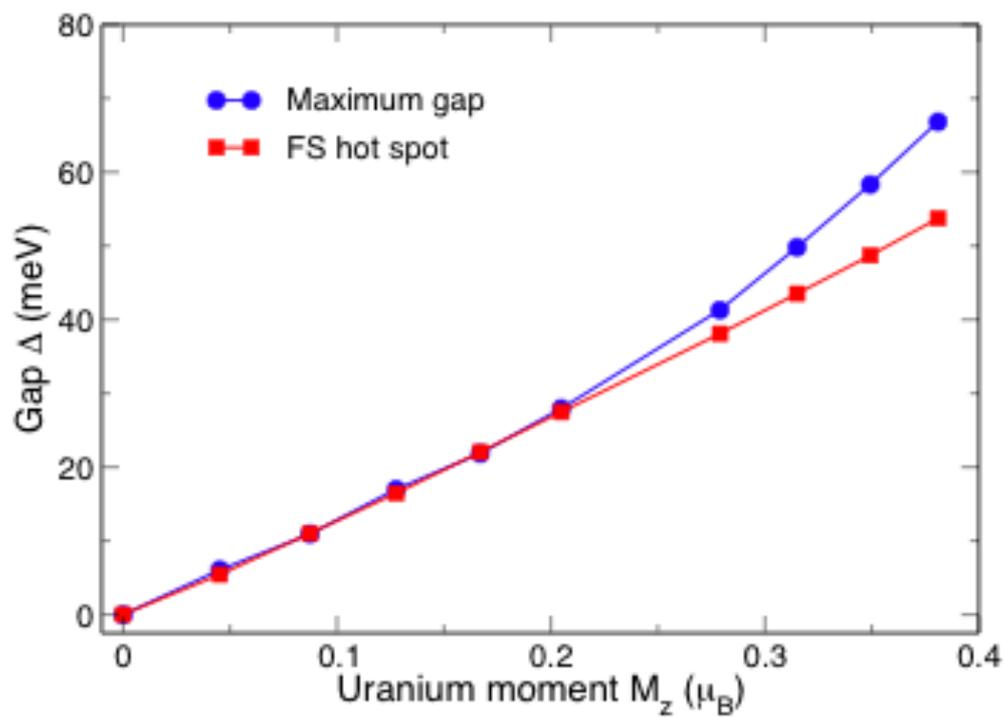

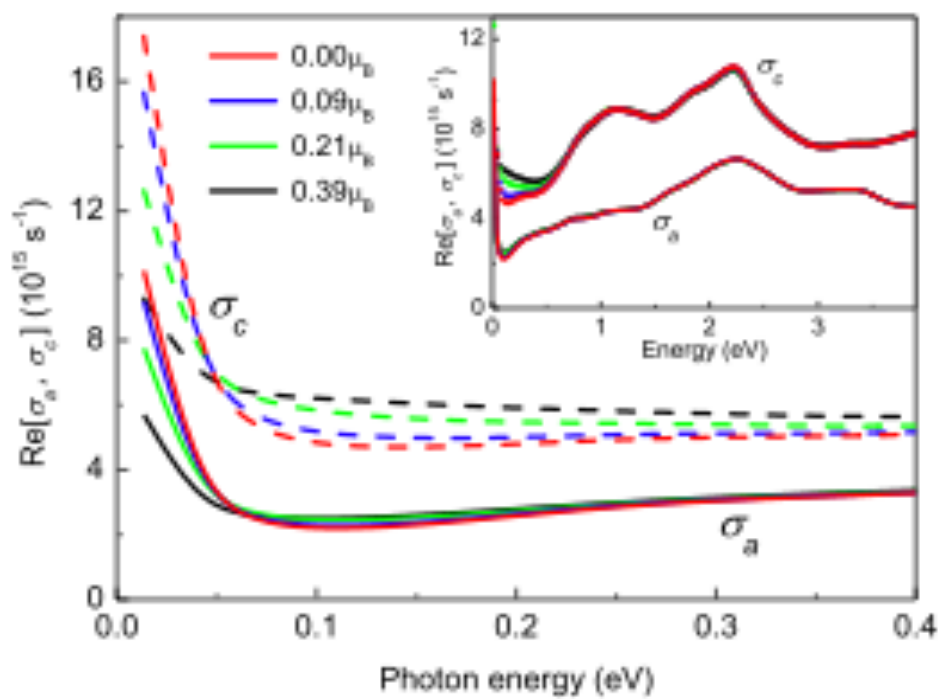